\RequirePackage[mathlines]{lineno}
\documentclass[aps,prd,twocolumn,showpacs,amsmath,amssymb]{revtex4-1}
\usepackage[T1]{fontenc}
\usepackage{amsmath}
\usepackage{graphicx}
\usepackage{appendix}
\usepackage{subfigure}
\usepackage{epstopdf}
\usepackage{color}
\usepackage{adjustbox}
\usepackage{multirow} 
\usepackage{setspace}
\usepackage{overpic}
\usepackage{tabularx}
\usepackage{amssymb,float}
\usepackage[driverfallback=dvipdfmx, bookmarksnumbered, pdfstartview=FitH,colorlinks,urlcolor=blue, citecolor=blue,linkcolor=blue] {hyperref}
\usepackage{lineno}
\usepackage{bm}
\usepackage{rotating}
\usepackage[utf8]{inputenc}

\let\oldequation\equation
\let\oldendequation\endequation

\renewenvironment{equation}
  {\linenomathNonumbers\oldequation}
  {\oldendequation\endlinenomath}

\newcommand{\BESIIIorcid}[1]{\href{https://orcid.org/#1}{\hspace*{0.1em}\raisebox{-0.45ex}{\includegraphics[width=1em]{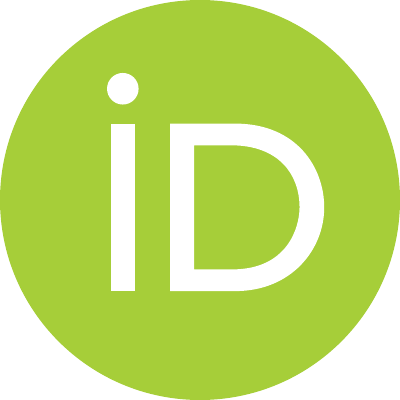}}}} 
\begin{document}

\title{\bf \boldmath
Search for the doubly Cabibbo-suppressed decays $D^0\to K^+\pi^-\eta^\prime$ and $D^+\to K^+\pi^0\eta^\prime$
}

\author{
	M.~Ablikim$^{1}$\BESIIIorcid{0000-0002-3935-619X},
M.~N.~Achasov$^{4,c}$\BESIIIorcid{0000-0002-9400-8622},
P.~Adlarson$^{88}$\BESIIIorcid{0000-0001-6280-3851},
X.~C.~Ai$^{95}$\BESIIIorcid{0000-0003-3856-2415},
C.~S.~Akondi$^{33A,33B}$\BESIIIorcid{0000-0001-6303-5217},
R.~Aliberti$^{42}$\BESIIIorcid{0000-0003-3500-4012},
A.~Amoroso$^{87A,87C}$\BESIIIorcid{0000-0002-3095-8610},
L.~P.~An$^{54,h}$\BESIIIorcid{0000-0002-3274-5627},
Q.~An$^{83,70,\dagger}$,
M.~S.~Anderson$^{42}$\BESIIIorcid{0009-0008-1550-2632}
% ...
\\
\vspace{0.2cm}
(BESIII Collaboration)\\
\vspace{0.2cm} {\it
$^{1}$ Institute of High Energy Physics, Beijing 100049, People's Republic of China\\
% ...
$^{4}$ Budker Institute of Nuclear Physics SB RAS (BINP), Novosibirsk 630090, Russia\\
% ...
$^{33}$ INFN Sezione di Ferrara, (A)INFN Sezione di Ferrara, I-44122, Ferrara, Italy; (B)University of Ferrara, I-44122, Ferrara, Italy\\
% ...
$^{42}$ Johannes Gutenberg University of Mainz, Johann-Joachim-Becher-Weg 45, D-55099 Mainz, Germany\\
% ...
$^{54}$ Peking University, Beijing 100871, People's Republic of China\\
% ...
$^{70}$ State Key Laboratory of Particle Detection and Electronics, Beijing 100049, Hefei 230026, People's Republic of China\\
% ...
$^{83}$ University of Science and Technology of China, Hefei 230026, People's Republic of China\\
% ...
$^{87}$ University of Turin and INFN, (A)University of Turin, I-10125, Turin, Italy; (B)University of Eastern Piedmont, I-15121, Alessandria, Italy; (C)INFN, I-10125, Turin, Italy\\
$^{88}$ Uppsala University, Box 516, SE-75120 Uppsala, Sweden\\
% ...
$^{95}$ Zhengzhou University, Zhengzhou 450001, People's Republic of China\\

\vspace{0.2cm}
$^{\dagger}$ Deceased\\
% ...
$^{c}$ Also at the Novosibirsk State University, Novosibirsk, 630090, Russia\\
% ...
$^{h}$ Also at State Key Laboratory of Nuclear Physics and Technology, Peking University, Beijing 100871, People's Republic of China\\
% ...
}

}

\begin{abstract}
  We present the first search for the doubly Cabibbo-suppressed decays $D^0\to K^+\pi^-\eta^\prime$ and $D^+\to K^+\pi^0\eta^\prime$ using an $e^+e^-$ collision data sample corresponding to an integrated luminosity of  20.3~fb$^{-1}$, collected at a center-of-mass energy of 3.773 GeV with the Beijing Spectrometer III (BESIII) detector at the Beijing Electron–Positron Collider II (BEPCII). No significant signals are observed, and the upper limits on their decay branching fractions are set to be $3.0\times 10^{-5}$ and $2.1\times 10^{-5}$ at the 90\% confidence level, respectively.
  By combining these results with the world-average  branching fractions of the corresponding Cabibbo-favored decays,
  upper limits at the 90\% confidence level are obtained on the ratios of doubly Cabibbo-suppressed to Cabibbo-favored branching fractions.
    The limits are determined to be $1.6\times \tan^4\theta_C$ and $3.7\times \tan^4\theta_C$ for $D^0\to K^+\pi^-\eta^\prime$ and $D^+\to K^+\pi^0\eta^\prime$, respectively, where $\theta_C$ denotes the Cabibbo mixing angle. 
\end{abstract}
\maketitle
\oddsidemargin  -0.2cm
\evensidemargin -0.2cm
\section{Introduction}

\begin{figure*}
	\includegraphics[width=0.32\linewidth]{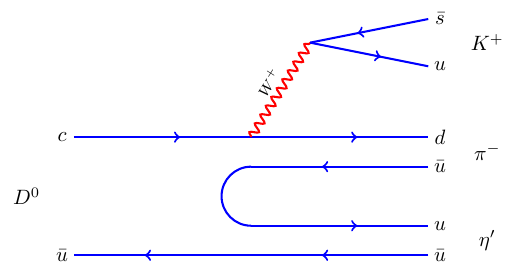}
	\includegraphics[width=0.32\linewidth]{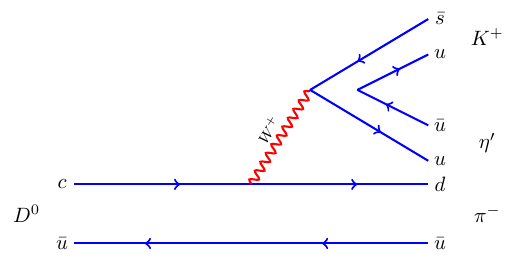} \\
	\includegraphics[width=0.32\linewidth]{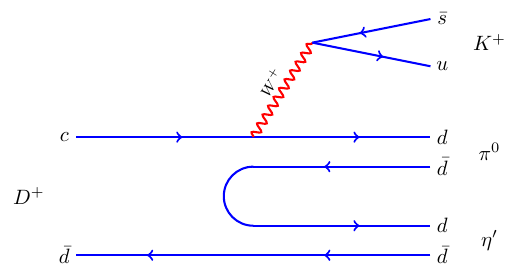}
	\includegraphics[width=0.32\linewidth]{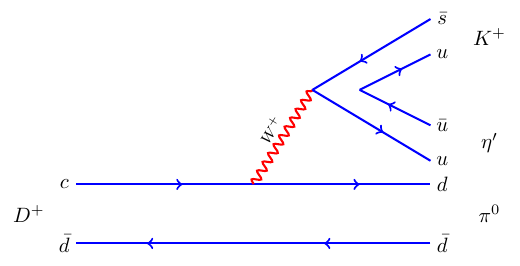}
	\includegraphics[width=0.32\linewidth]{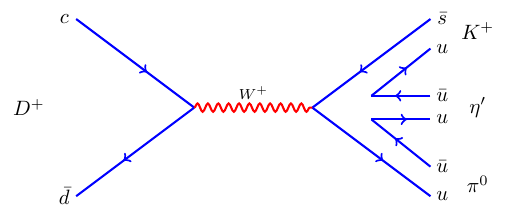}
	\vspace{6pt}
	\caption{Feynman diagrams of the DCS decays  (top row) $D^0\to K^+\pi^-\eta^\prime$ and (bottom row) $D^+\to K^+\pi^0\eta^\prime$.}
	\label{feynman}
\end{figure*}

Investigations of doubly Cabibbo-suppressed (DCS) decays of charmed mesons are
important for understanding the decay dynamics of charm quarks.
It is naively expected that the branching fraction (BF) of the DCS $D^{0(+)}$ decay is $(0.5-2.0)$ ${\rm\tan}^4\theta_C$ times that of its corresponding Cabibbo-favored (CF)
decay~\cite{Lipkin:2002za,Cheng:2010ry}, where ${\rm\theta}_{C}$ denotes the Cabibbo mixing angle~\cite{Cabibbo:1963yz}.
All known DCS $D^{0(+)}$ decays are roughly consistent with this expectation~\cite{Charmreview,ParticleDataGroup:2024}, except for $D^+\to K^+\pi^+\pi^-\pi^0$.
This decay was observed for the first time by BESIII~\cite{BESIII:2020kpipipi0,BESIII:2021kpipipi0} and later confirmed by Belle~\cite{BELLE:2023kpipipi0}.
The world-average BF of this decay is $(1.21 \pm 0.09)\times 10^{-3}$, corresponding to a DCS-to-CF branching-fraction ratio of $(6.70\pm0.49)\tan^4\theta_{C}$, which is substantially larger than the naive expectation.
The BFs of the DCS decays $D^0\to K^{*0}\eta^\prime$ and $D^+\to K^{*+}\eta^\prime$, which may contribute to the final states $D^0\to K^+\pi^-\eta^\prime$ and $D^+\to K^+\pi^0\eta^\prime$, are predicted within the pole model to be $(0.40\pm0.10)\times 10^{-6}$ and $(0.20\pm0.70)\times 10^{-5}$~\cite{theory1}. In 2018, BESIII reported the first observation of the CF decays $D^0\to K^-\pi^+\eta^\prime$ and $D^+\to K_S^0\pi^+\eta^\prime$, and determined their decay BFs to be $(6.43\pm0.34)\times 10^{-3}$ and $(1.90\pm0.21)\times 10^{-3}$~\cite{BESIII:2018wnc}. The DCS decays $D^0\to K^+\pi^-\eta^\prime$ and $D^+\to K^+\pi^0\eta^\prime$ can  proceed through the Feynman diagrams shown in Fig.~\ref{feynman}. To test the theoretical predictions on DCS-to-CF ratios, it is urgent for us to search for the corresponding DCS decay modes $D^0\to K^+\pi^-\eta^\prime$ and $D^+\to K^+\pi^0\eta^\prime$.
Charge conjugate decays are always implied throughout this paper.

In this paper, we report the first search for the DCS decays
$D^0\to K^+\pi^-\eta^\prime$ and $D^+\to K^+\pi^0\eta^\prime$,
based on $e^+e^-$ collision data corresponding to an integrated luminosity of  20.3 fb$^{-1}$~\cite{BESIII:DDluminosity},
 taken at a center-of-mass energy of $\sqrt s=3.773$~GeV with the BESIII detector during 2010-2011 and 2022-2024. Here, the double-tag (DT) method~\cite{MARK-III:1985hbd,MARK-III:1987jsm} is employed to reconstruct $D^0\bar{D}^0$ or $D^+D^-$ pairs. The decay modes used to reconstruct the $\bar{D}^0$ or $D^-$ meson are referred to as single-tag (ST) modes. Signal candidates are then identified in the system recoiling against the ST $\bar{D}^0$ or $D^-$, corresponding to the accompanying $D^0$ or $D^+$.
In particularly, the CF and DCS $D^0$ decays to the same final state are experimentally indistinguishable, leading to significantly larger backgrounds when studying DCS $D^0$ decays with conventional hadronic ST modes. Such analyses also require nontrivial corrections for quantum-correlation effects,
especially for decay modes whose strong-phase difference parameters are not experimentally known~\cite{ParticleDataGroup:2024}.
Therefore, semileptonic (SL) decays are employed as ST modes to  investigate  DCS $D^0$ decays, for which the effect of $D^0\bar D^0$ mixing is negligible. 
By combining our results with the world-average BFs of the corresponding CF decays, 
the ratios of the DCS-to-CF branching fractions, $\mathcal B_{\rm DCS}/\mathcal B_{\rm CF}$, are determined and expressed in units of $\tan^{4} \theta_C$.

\section{Data and Monte Carlo Generation}

The BESIII detector~\cite{BESIII:2009fln} records symmetric $e^+e^-$ collisions provided by the BEPCII storage ring~\cite{Yu:2016cof}
in the center-of-mass energy range from 1.84 to 4.95~GeV,
with a peak luminosity of $1.1 \times 10^{33}\;\text{cm}^{-2}\text{s}^{-1}$
achieved at $\sqrt{s} = 3.773\;\text{GeV}$.

The cylindrical core of the BESIII detector covers 93\% of the full solid angle and consists of a helium-based
multilayer drift chamber~(MDC), a plastic scintillator time-of-flight
system~(TOF), and a CsI(Tl) electromagnetic calorimeter~(EMC),
which are all enclosed in a superconducting solenoidal magnet
providing a 1.0~T magnetic field.
The solenoid is supported by an
octagonal flux-return yoke with resistive plate counter muon
identification modules interleaved with steel.
The charged-particle momentum resolution at $1~{\rm GeV}/c$ is
$0.5\%$, and the
${\rm d}E/{\rm d}x$
resolution is $6\%$ for electrons
from Bhabha scattering. The EMC measures photon energies with a
resolution of $2.5\%$ ($5\%$) at $1$~GeV in the barrel (end-cap)
region. The time resolution in the TOF barrel region is 68~ps, while
that in the end-cap region was 110~ps.
The end-cap TOF system was upgraded in 2015 using multi-gap resistive plate chamber technology, providing a time resolution of 60~ps~\cite{Tof1,Tof2,Tof3}, which benefits 86\% of the data used in this analysis.

Simulated samples produced with the {\sc geant4}-based~\cite{GEANT4:2002zbu} Monte Carlo (MC) package, which
includes the geometric description of the BESIII detector and the
detector response, are used to determine the detection efficiency
and to estimate the backgrounds. The simulation includes the beam-energy spread and initial-state radiation in the $e^+e^-$
annihilations modeled with the generator {\sc kkmc}~\cite{Jadach:2000ir,Jadach:1999vf}.
The inclusive MC samples are about 40 times the luminosity of the data sample, which include the production of $D\bar{D}$ pairs,
the non-$D\bar{D}$ decays of the $\psi(3770)$, the initial-state radiation
production of the $J/\psi$ and $\psi(3686)$ states, and the
continuum processes.
The known decay modes are modelled with {\sc
    evtgen}~\cite{Ping:2008zz,Lange:2001uf} using the BFs taken from the
Particle Data Group (PDG)~\cite{ParticleDataGroup:2024}, and the remaining unknown decays of the charmonium states are
modeled by {\sc lundcharm}~\cite{Chen:2000tv}. Final-state radiation is incorporated using the {\sc photos} package~\cite{Richter-Was:1992hxq}.

To obtain accurate estimates of the signal efficiencies and signal shapes, we generate $2.0\times10^6$ MC events for each of the channels  $D^0\to K^+\pi^-\eta'_{\pi^+\pi^-\eta}$, $D^0\to K^+\pi^-\eta'_{\pi^+\pi^-\gamma}$ and $D^+\to K^+\pi^0\eta'_{\pi^+\pi^-\eta}$. The DCS $D$ decays are simulated with a phase-space model.
The $\bar D^0 \to K^+e^- \bar\nu_e$ decay is simulated with
a specific two-parameter series expansion model~\cite{Becirevic:1999kt,BESIII:2015tql,Tchikilev:1999vpw}. The $D^-\to K^+\pi^-\pi^-$ decay is simulated with the Dalitz model. The $\eta^\prime\to \pi^+\pi^-\eta$ and $\eta^\prime\to \pi^+\pi^-\gamma$ decays are simulated with  generators based on amplitude analysis results~\cite{Gen:pipieta,Gen:pipigamma}.

\section{Event selection}
\label{eventselection}
Charged tracks detected in the MDC are required to be within a polar angle ($\theta$) range of $|\rm{cos\theta}|<0.93$, where $\theta$ is defined with respect to the $z$-axis, which is the symmetry axis of the MDC.
For charged tracks not originating from $K_S^0$ decays, the distance of closest approach to the interaction point (IP)  must be less than 10\,cm along the $z$-axis, $|V_{z}|$,  and less than 1\,cm in the transverse plane, $|V_{xy}|$.
Particle identification~(PID) for charged tracks combines measurements of the energy deposited in the MDC~(d$E$/d$x$) and the flight time in the TOF to form likelihoods $CL_h~(h=K,\pi)$ for each hadron $h$ hypothesis.
Charged tracks with  $CL_K>CL_\pi$ and $CL_\pi>CL_K$ are assigned as charged kaons and pions, respectively.

Each $K_{S}^0$ candidate is reconstructed from two oppositely charged tracks satisfying $|V_{z}|<$ 20~cm.
The two charged tracks are assigned as $\pi^+\pi^-$ without imposing further PID criteria. They are constrained to
originate from a common vertex and are required to have an invariant mass within $M_{\pi^{+}\pi^{-}} \in$ (0.487, 0.511)~GeV$/c^{2}$. The
decay length of the $K^0_S$ candidate is required to be greater than
twice the vertex resolution away from the IP.	
The quality of the vertex fit is required to be  $\chi^2<200$.

Photon candidates are selected using isolated showers in the EMC.
The deposited energy is required to be greater than $25~\mathrm{MeV}$ in the barrel $(|\rm{cos\theta}|<0.8)$ and $50~\mathrm{MeV}$ in the end cap $(0.86<|\rm{cos\theta}|<0.92)$ regions.
To exclude showers that originate from charged tracks, the angle subtended by the EMC shower and the position of the closest charged track at the EMC must be greater than 10 degrees as measured from the IP. 
To suppress electronic noise and showers unrelated to the event, the difference between the EMC time and the event start time is required to be within 
[0, 700]\,ns.
The $\pi^0$ and $\eta$ candidates are formed from photon pairs with invariant mass within $(0.115,\,0.150)$ GeV$/c^{2}$ and $(0.505,\,0.575)$ GeV$/c^{2}$, respectively.
To improve the resolution, a kinematic fit constraining the $\gamma\gamma$ invariant mass to the nominal $\pi^{0}$ or $\eta$ mass~\cite{ParticleDataGroup:2024} is imposed on the selected photon pairs.
The $\chi^2$ of the kinematic fit is required to be less than 50.
The four-momentum of the $\pi^0$ or $\eta$ candidate, as updated by the kinematic fit, is used in the subsequent analysis.

The $\eta^\prime$ candidates are reconstructed via the $\eta^\prime\to \pi^+\pi^-\eta$ and $\eta^\prime\to \rho^0\gamma$ decays, where the $\rho^0$ candidates are required to satisfy $M_{\pi^+\pi^-}\in$  (0.62, 0.92) GeV/$c^2$;
the signal region is defined as the $\pm 3\sigma$ interval around the fitted $\eta^\prime$ mass, in which the $\eta^\prime$ signal shape is described by a double-Gaussian function for $M_{\pi^+\pi^-\eta}$ and a Crystal-ball function for $M_{\rho^0\gamma}$, and the background is described by a third-order Chebyshev polynomial function. The $\eta^\prime$ candidates reconstructed via $\eta^\prime\to \pi^+\pi^-\eta$ and $\eta^\prime\to \rho^0\gamma$ are required to satisfy $M_{\pi^+\pi^-\eta}\in$ (0.943, 0.973) GeV/$c^2$ and $M_{\rho^0 \gamma}\in$(0.938, 0.978) GeV/$c^2$, respectively.
\section{Double-tag events}
At $\sqrt s=3.773$~GeV, $D^0\bar D^0$ pairs are produced copiously without accompanying hadrons.
This clean production environment enables the use of the double-tag technique to study the $D$-meson decays.
\subsection{Selection of $D^0\to K^+\pi^-\eta^\prime$}

In this section, DT events refer to those in which the DCS $D^0$ decay is reconstructed on the side recoiling against the SL decay $\bar D^0\to K^+ e^-\bar \nu_e$.
The BF of the $D^0\to K^+\pi^-\eta^\prime$ decay is determined by
\begin{equation}\label{equ:br}
  {\mathcal B}_{{\rm sig}} = \frac{N_{\rm DT}} {2\cdot N_{D^0\bar D^0}\cdot
    \epsilon_{\rm DT}\cdot  {\mathcal B}_{\rm SL}},
\end{equation}
where $N_{D^0\bar D^0}=(73.29\pm0.84) \times 10^6$
is the total number of $D^0\bar D^0$ pairs in the data sample, calculated
from the $e^+e^-\to D^0\bar D^0$ production cross section at $\sqrt s=3.773$ GeV~\cite{BESIII:DDluminosity} and the integrated luminosity of the data sample~\cite{BESIII:DDluminosity};
$N_{\rm DT}$ is the number of the DT events observed in the data sample;
$\epsilon_{\rm DT}$ is the efficiency of reconstructing DT events estimated from MC simulation,
and
${\mathcal B}_{\rm SL}$ is the BF of the $\bar D^0\to K^+ e^-\bar \nu_e$ decay quoted from the PDG~\cite{ParticleDataGroup:2024}.

In the selection of the  $D^0\to K^+\pi^-\eta^\prime(\pi^+\pi^-\eta)$ and $D^0\to K^+\pi^-\eta^\prime(\rho^0\gamma)$ candidates,
the $\pi^+\pi^-$ combinations are required to have an invariant mass outside the window $|M_{\rm\pi^+\pi^-}-0.4977|<0.03$~GeV/$c^2$, to reject the dominant peaking background from the $K_S^0\eta^\prime$ process.
Additionally, the number of extra charged tracks ($N_{\rm extra}^{\rm charge}$) is required to be zero.
The charged kaons from the SL $\bar D^0$ decays are required to satisfy the same PID criteria as those from hadronic $D^0$ decays, and
the lepton candidate is required to have the opposite charge to the kaon from the SL $\bar D^0$ decays.
Electron PID is based on the combined information from the d$E$/d$x$, TOF, and EMC measurements, from which the confidence levels under the electron, pion, and kaon hypotheses ($\mathit{CL}_e$, $\mathit{CL}_{\pi}$, and $\mathit{CL}_{K}$) are calculated.
Electron candidates are required to satisfy $\mathit{CL}_e>0.001$ and $\mathit{CL}_e/(\mathit{CL}_e+\mathit{CL}_\pi+\mathit{CL}_K)>0.8$.
In addition, the ratio of the energy deposited in the EMC to the track momentum, $E/p$, is required to be greater than 0.8, where $E$ is the energy deposited in the EMC and $p$ is the track momentum.

To suppress background from hadronic decays in which a hadron is misidentified as an electron,
the invariant mass of the $K^+e^-$ pair, $M_{K^+e^-}$, is required to be less than 1.8~GeV/$c^2$.
Furthermore, the maximum energy of extra photons not used in the tag selection, $E^{\rm max}_{\rm extra,\gamma}$, is required to be less than 0.2~GeV, and no extra good $\pi^0$ candidate is allowed ($N_{\rm extra}^{\pi^0}=0$).

The signal candidates for $D^0\to K^+\pi^-\eta^\prime$ are identified with two variables: the energy difference
$\Delta E_{\rm sig} = E_{D^0} - E_{\rm beam}$
and the beam-constrained mass
$M^{\rm sig}_{\rm BC} = \sqrt{E^{2}_{\rm beam}-|\vec{p}_{D^0}|^{2}}.$
Here, $E_{\rm beam}$ is the beam energy, $\vec{p}_{D^0}$ and $E_{D^0}$ are the momentum and energy of the $D^0$ candidate in the $e^+e^-$ rest frame, respectively.
If multiple candidates are found on the hadronic side, only the one with the smallest $|\Delta E_{\rm sig}|$ is retained.
The correctly reconstructed $D^0$ candidates concentrate around zero in the $\Delta E_{\rm sig}$ distribution
and around the  $D^0$ mass in the $M^{\rm sig}_{\rm BC}$ distribution. 
The selected candidates are required to satisfy $\Delta E_{\rm sig}\in(-0.026, 0.024)$~GeV and $\Delta E_{\rm sig}\in(-0.033, 0.027)$~GeV for $D^0\to K^+\pi^-\eta^\prime$ with $\eta^\prime\to\pi^+\pi^-\eta$ and $\eta^\prime\to\rho\gamma$, respectively.

After reconstructing the signal-side $D^0$ candidates, $\bar D^0\to K^+e^-\bar{\nu}_e$ candidates are reconstructed from the remaining tracks that are not used in the signal-side reconstruction.
 The four-momenta of the photon(s) within 5$^\circ$ of the initial
electron direction are added to the electron four-momentum in order to partially compensate
the effects of the final-state radiation and bremsstrahlung (FSR recovery)~\cite{FSR}.

The SL decay $\bar{D}^0\to K^+e^-\bar{\nu}_e$ is identified using the kinematic variable
$ U_{\mathrm{miss}}= E_{\mathrm{miss}}-|\vec{p}_{\mathrm{miss}}|.$
Here, $E_{\mathrm{miss}}= E_{\mathrm{beam}}-E_{K^+}-E_{e^-}$ and $\vec{p}_{\mathrm{miss}}=
  \vec{p}_{\bar D^0}-\vec{p}_{K^+}-\vec{p}_{e^-}$ are, respectively, the missing energy and momentum of the DT event in the $e^+e^-$ center-of-mass system, in
which $E_{K^+}$ and $\vec{p}_{K^+}$ are the energy and momentum of the $K^+$, and
$E_{e^-}$ and $\vec{p}_{e^-}$ are the energy and momentum of the $e^-$. The
$U_{\mathrm{miss}}$ resolution is improved by constraining the $D^0$ energy to the beam energy and $\vec{p}_{\bar D^0} = {-\hat{p}_{D^0}}\cdot\sqrt{E_{\mathrm{beam}}^{2}-M_{D^0}^{2}}$, where $\hat{p}_{D^0}$ is the unit vector in the momentum direction of the $D^0$ and $M_{D^0}$ is the $D^0$ nominal mass~\cite{ParticleDataGroup:2024}.

\begin{figure}[htbp]
	\centering
	\includegraphics[width=0.45\textwidth]{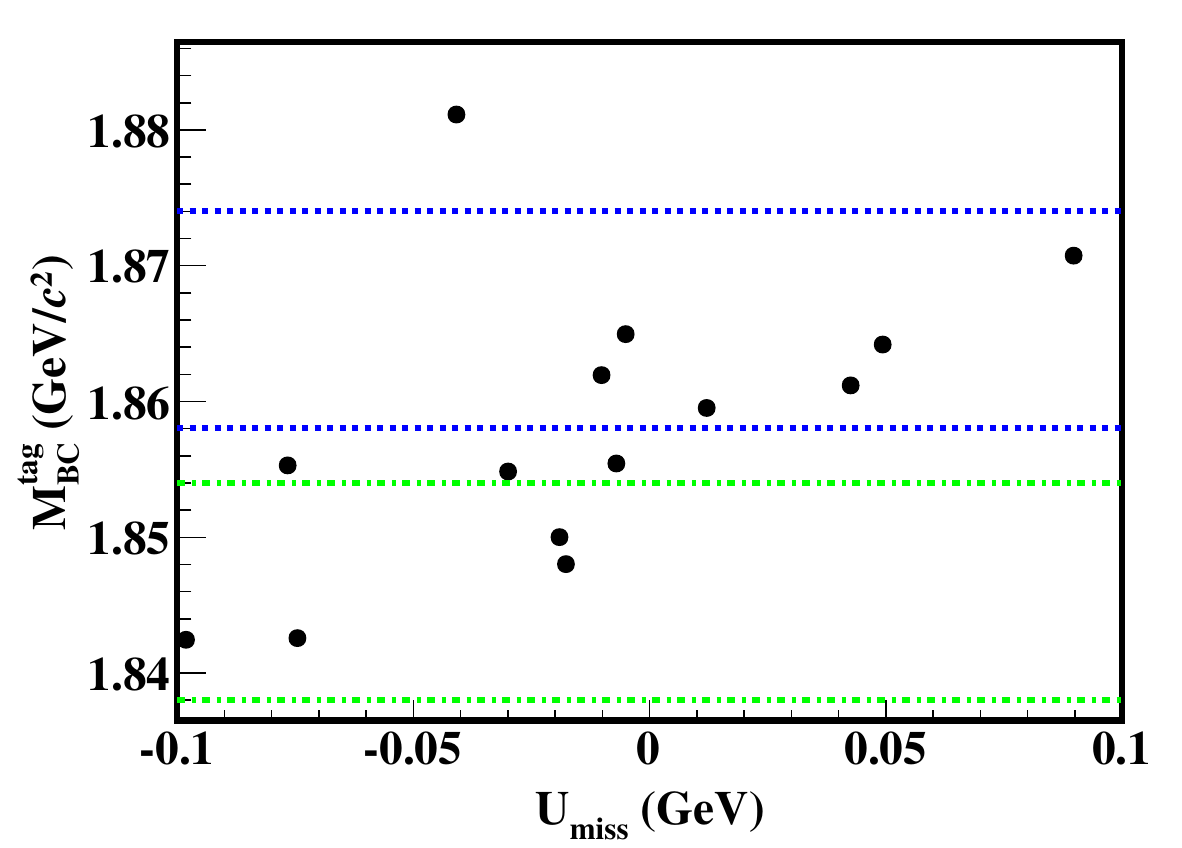}
	\vspace{10pt}
	\caption{Distribution of $M_{\rm BC}^{\rm tag}$ versus $U_{\rm miss}$ for $D^0 \to K^+\pi^-\eta^\prime$ candidates in data. The blue dashed lines show the signal region. The green dot-dashed lines show the sideband region.}
	\label{fig:M_BC_M2}
\end{figure}

Figure \ref{fig:M_BC_M2} shows the $M_{\rm BC}^{\rm tag}$ versus $U_{\rm miss}$ distribution of the accepted DT candidate events in data, where $\eta^\prime\to \pi^+\pi^-\eta$ and $\eta^\prime\to \rho\gamma$ are both included. 
There are 6 and 4 candidate events in the signal and sideband regions in data, respectively.
The signal and sideband regions are defined as $M_{\rm BC}\in(1.858, 1.874)$~GeV/$c^2$ and $M_{\rm BC}\in(1.838, 1.854)$~GeV/$c^2$, respectively.
The detection efficiencies $\epsilon_{{\rm DT}}$ are estimated with signal MC samples
to be $(5.84\pm0.02)\%$ and $(8.68\pm0.02)\%$ for $D^0\to K^+\pi^-\eta^\prime$ with $\eta^\prime\to \pi^+\pi^-\eta$ and $\eta^\prime\to \rho^0\gamma$, respectively. Here the BFs of $\eta^\prime$ decays are not included.

\subsection{Selection of $D^+\to K^+\pi^0\eta^\prime$}
\begin{figure}[htbp]\centering
  \includegraphics[width=0.95\linewidth]{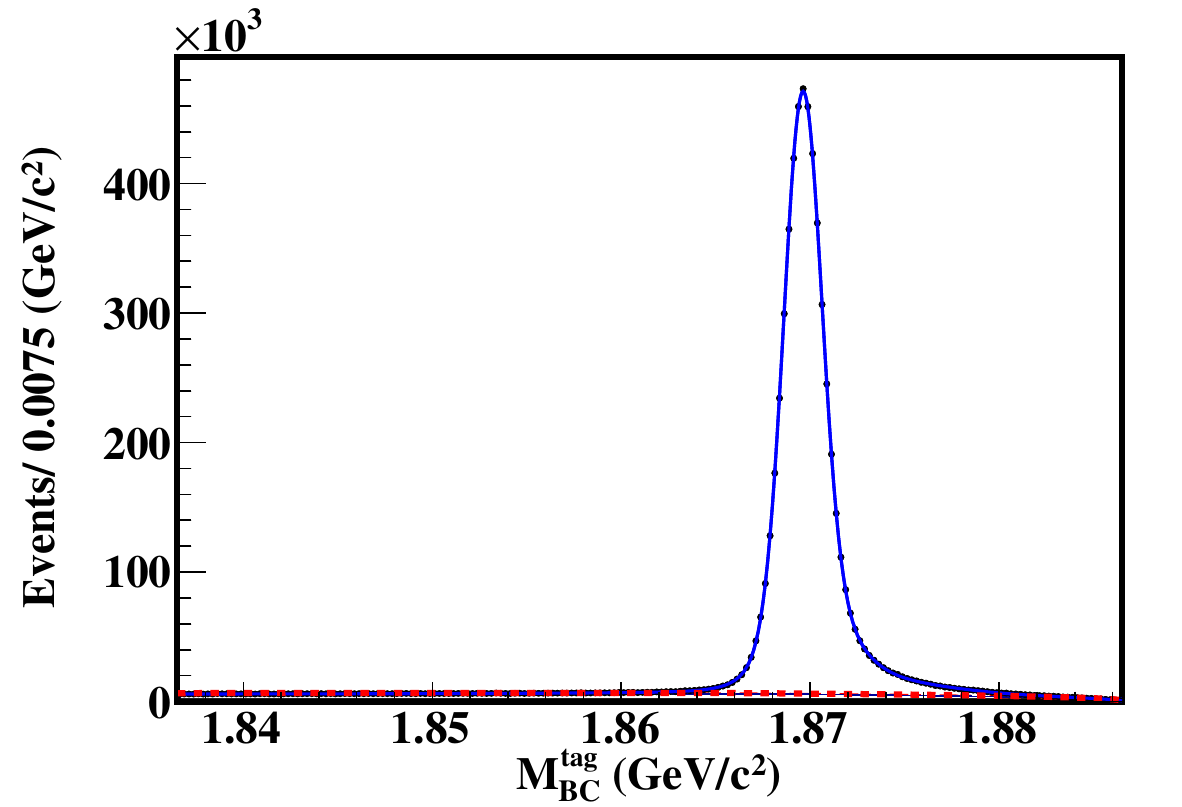}
  \vspace{10pt}
  \caption{
    Fit to the $M^{\rm tag}_{\rm BC}$ distribution of the ST $D^-$ candidates for $D^-\to K^+\pi^-\pi^-$.
    The dots with error bars are data, the blue curve is the best fit, and the red dashed curve describes the background.
  }\label{fig:datafit_Massbc}
\end{figure}

The ST $D^-$ candidates are selected by reconstructing a $D^-$ in the hadronic decay mode 
$D^- \to K^{+}\pi^{-}\pi^{-}$. Figure~\ref{fig:datafit_Massbc} shows the fit to the $M^{\rm tag}_{\rm BC}$ distribution of the ST $D^-$ candidates for $D^-\to K^+\pi^-\pi^-$. This tag mode is chosen from the six commonly used $D^-$ ST modes, $D^-\to K^+\pi^-\pi^-$, $K_S^0\pi^-$, $K^+\pi^-\pi^-\pi^0$, $K_S^0\pi^-\pi^0$, $K_S^0\pi^+\pi^-\pi^-$ and $K^+K^-\pi^-$, which are widely used in studies of $D^+$ decays, by maximizing the Punzi figure of merit, $\epsilon/(1.5+\sqrt{B})$~\cite{punzi}. Here, $\epsilon$ is the efficiency for the signal decay, and $B$ is the scaled background yield estimated by the inclusive MC sample.
The events in which a signal candidate is reconstructed in the presence of a ST $D^-$ meson
are referred to as DT events.
The BF of $D^+\to K^+\pi^0\eta^\prime$ is determined by
\begin{equation}
  \label{eq:br}
  {\mathcal B}_{{\rm sig}} = N_{\rm DT}/(N_{\rm ST}\cdot\epsilon_{{\rm sig}}),
\end{equation}
where
$N_{\rm ST}$ is the yield of the ST candidates in the data sample.
The efficiency $\epsilon_{{\rm sig}}$ for detecting the signal $D^+$ decay is given as
\begin{equation}
  \epsilon_{\text {sig}}=\epsilon_{\mathrm{DT}} / \epsilon_{\mathrm{ST}},
\end{equation}
where $\epsilon_{\mathrm{ST}}$ is the ST reconstruction efficiency. 

The selection criteria of $K^\pm$, $\pi^\pm$, $\pi^0$ and $\eta$
have been discussed in Section~\ref{eventselection}.
The ST $D^-$ mesons are identified using two variables: the energy difference $\Delta E_{\rm tag} = E_{D^-} - E_{\rm beam}$
and the beam-constrained mass $M_{\rm BC}^{\rm tag} = \sqrt{E^{2}_{\rm beam}-|\vec{p}_{D^-}|^{2}}$. Here, $\vec{p}_{D^-}$ and $E_{D^-}$ are the momentum and energy of the $D^-$ in the rest frame of the $e^+e^-$ system, respectively.
For each ST mode, if multiple candidates are found in an event, only the one with the smallest $|\Delta E_{\rm tag}|$ is retained.
The $\Delta E_{\rm tag}$ of ST $D^-$ candidates is required to be within $(-0.025, 0.021)$~GeV.

To extract the yield of ST $D^-$ candidates, a binned maximum-likelihood fit is performed on the $M_{\rm BC}^{\rm tag}$
distribution of the accepted ST candidates following Ref.~\cite{BESIII:2023exq}.
The $D^-$ signal is modeled by the MC-simulated shape convolved with
a double-Gaussian function that accounts for the resolution difference between data and MC simulation.
The shape of the combinatorial background  is described by an ARGUS function~\cite{ARGUS:1990hfq}.
The ST $D^-$ yield in data, $N_{\rm ST}$, is obtained to be $(5674.2\pm2.5)\times 10^{3}$. The ST efficiency for $D^-\to K^+\pi^-\pi^-$ is estimated by analyzing the inclusive MC sample to be $(52.40\pm 0.01)$\%.

\begin{figure}[htbp]
	\centering
	\includegraphics[width=0.46\textwidth]{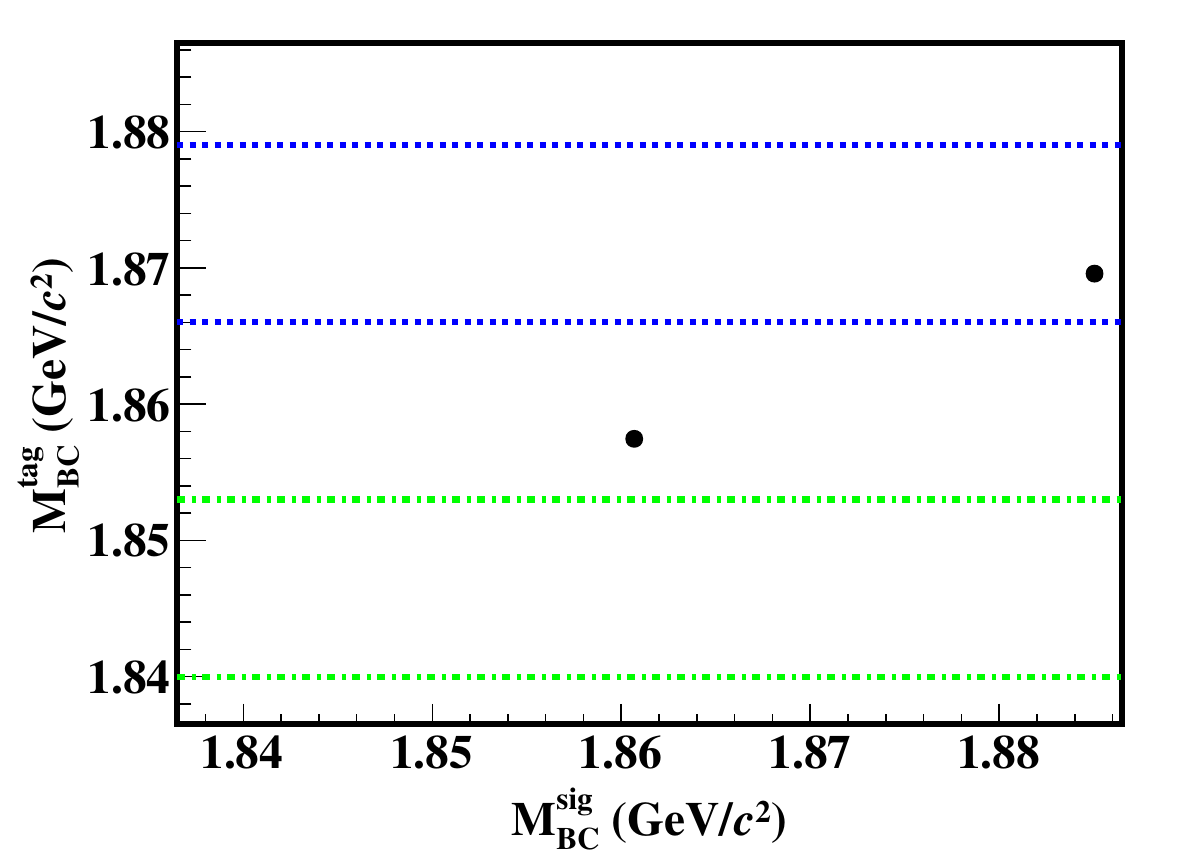}
	\vspace{10pt}
	\caption{Distribution of $M_{\rm BC}^{\rm tag}$ versus $M_{\rm BC}^{\rm sig}$  for $D^+ \to K^+\pi^0\eta^\prime$ candidates in data. The blue dashed lines show the signal region. The green dot-dashed lines show the sideband region.}
	\label{fig:M_BC_M22}
\end{figure}

Candidates for the signal decay $D^+\to K^+\pi^0\eta^\prime$ are reconstructed from the remaining tracks and showers recoiling against the ST $D^-$ candidates. To obtain a better significance, the $\eta^\prime$ candidates are only reconstructed via the $\eta^\prime\to \pi^+\pi^-\eta$ decay. The $\eta^\prime$ selection criteria  are the same as those described in Section~\ref{eventselection}.

The $D^+\to K^+\pi^0\eta^\prime$ signal candidates are identified using the energy difference $\Delta E_{\rm sig}$ and the beam-constrained mass $M_{\rm BC}^{\rm sig}$, calculated similarly to that of the tag side.
If there are multiple candidates in an event, only the one giving the minimum $|\Delta E_{\rm sig}|$ is chosen.
The $\Delta E_{\rm sig}$ of selected candidates is required to be within $(-0.035, 0.030)$ GeV.

Figure \ref{fig:M_BC_M22} shows the $M_{\rm BC}^{\rm tag}$ versus $M_{\rm BC}^{\rm sig}$ distribution of the accepted DT candidate events in data; there are 1 and 0 candidate events in the signal and sideband regions, respectively. The signal and sideband regions are defined as $M_{\rm BC}\in(1.866, 1.879)$~GeV/$c^2$ and $M_{\rm BC}\in(1.840, 1.853)$~GeV/$c^2$, respectively.  The detection efficiency $\epsilon_{{\rm DT}}$ is estimated from signal MC samples
to be $(4.35\pm0.01)\%$ for $D^+\to K^+\pi^0\eta^\prime$. Here the BF of $\eta^\prime$ decay is not included.

\section{Systematic uncertainties}

The systematic uncertainties in the measurements of the BFs of the DCS decays of $D^0\to K^+\pi^-\eta^\prime$ and $D^+\to K^+\pi^0\eta^\prime$ are discussed below.

The $e^-$ tracking and PID efficiencies
are studied by using a control sample of $e^+e^-\to\gamma e^+ e^-$ events~\cite{BESIII:klnu}.
The systematic uncertainty in the $e^-$ tracking and PID is assigned as 0.5\% per $e^-$.
The $K^\pm$ and $\pi^\pm$ tracking and PID efficiencies are investigated with the DT hadronic $D\bar D$ events, including $D^0 \to K^-\pi^+$, $K^-\pi^+\pi^0$, $K^-\pi^+\pi^+\pi^-$ versus~$\bar D^0 \to K^+\pi^-$, $K^+\pi^-\pi^0$, $K^+\pi^-\pi^-\pi^+$, and $D^+\to K^- \pi^+\pi^+$ versus~$D^-\to K^+\pi^-\pi^-$.
By weighting the data-MC difference at different momentum bins, the systematic uncertainty of the tracking or PID is assigned as 0.5\% per $K^\pm$ or $\pi^\pm$.

The $K^\pm/\pi^\pm$ misidentification efficiencies are also studied using control samples of $D^0\to K^-\pi^+$, $K^-\pi^+\pi^0$, $K^-\pi^+\pi^+\pi^-$ versus $\bar{D}^0\to K^+\pi^-$, $K^+\pi^-\pi^0$, $K^+\pi^-\pi^-\pi^+$, as well as $D^+\to K^-\pi^+\pi^+$ versus $D^-\to K^+\pi^-\pi^-$. By weighting the $K^\pm/\pi^\pm$ misidentification rates in different $p_K$ intervals, the corresponding systematic uncertainty is determined to be 0.5\% for both $D^0\bar{D}^0$ and $D^+D^-$ events.

The $\pi^0$ reconstruction efficiency is studied using DT hadronic $D\bar D$ events~\cite{BESIII:nkpi}. Due to the limited size of the $\eta$ control sample and its similar topology to the $\pi^0$ sample, the uncertainties associated with both $\pi^0$ and $\eta$ reconstruction are taken to be 2.0\%.

The $\eta^\prime$ mass-window requirement is studied using control samples of the CF decays $D^0\to K^-\pi^+\eta^\prime$ and $D^+\to K_S^0\pi^+\eta^\prime$. The differences in the efficiencies obtained using different $\eta^\prime$ mass-window selections are assigned as the corresponding systematic uncertainties.

The systematic uncertainties associated with the $\Delta E$ and $M_{\rm BC}$ requirements on the hadronic side are evaluated
 using control samples of the CF decays $D^0\to K^-\pi^+\eta^\prime$ and $D^+\to K_S^0\pi^+\eta^\prime$.
The
differences in the acceptance efficiencies between data and MC simulation are found to be 0.3\% and 0.6\% for the $D^0$ and $D^+$ decays, respectively, and are assigned as the corresponding systematic uncertainties.

The systematic uncertainty associated with the $M_{K^+e^-}$ requirement for $\bar{D}^0\to K^+e^-\bar{\nu}_e$ is estimated using $\bar{D}^0\to K^+e^-\bar{\nu}_e$ candidates tagged by CF decays, and is assigned to be 0.1\%~\cite{BESIII:DDluminosity}.

The systematic uncertainty associated with the $E^{\rm max}_{\rm extra\, \gamma}$, $N_{\rm extra}^{\rm charge}$ and $N_{\rm extra}^{\pi^0}$ requirements
is evaluated using a control sample of $\bar D^0\to K^+e^-\bar\nu_e$ candidates tagged by CF decays.
The difference in the acceptance efficiencies between data and MC simulation is taken as the systematic uncertainty.

The systematic uncertainties associated with the $U_{\rm miss}$ and $M_{\rm BC}$ fits are evaluated by varying the fit range and the smoothing parameter of the MC-simulated shape. The maximum differences in the fitted DT yields are taken as the corresponding systematic uncertainties. 

The systematic uncertainty due to the limited MC statistics is estimated as
$\frac{1}{\sqrt{N}} \sqrt{(1-\epsilon)/\epsilon}$,
where $\epsilon$ is the detection efficiency and $N$ is the total number of generated signal MC events.

The systematic uncertainties due to the MC model are evaluated by comparing the DT efficiencies obtained from signal MC events of the $D^{0(+)}\to K^+\pi^{-(0)}\eta^\prime$ and $D^{0(+)}\to K^{*0(+)}\eta^\prime$ decays. The maximum differences in the DT efficiency are taken as the corresponding systematic uncertainties.

The systematic uncertainty associated with the FSR recovery for electron candidates is assigned to be 0.3\%, according to Ref.~\cite{BESIII:2015tql}.

The systematic uncertainty in the total number of $D^0\bar D^0$ pairs
in data is 1.2\%~\cite{BESIII:DDluminosity}.
The systematic uncertainty in the ST yield in data
is 0.3\%~\cite{BESIII:DDluminosity}.

The uncertainties in the quoted BFs of $D^0\to K^-e^+\nu_{e}, \eta^\prime\to\pi^+\pi^-\eta, \eta^\prime\to\rho^0\gamma, \eta\to\gamma\gamma$ and $\pi^0\to \gamma \gamma$ are $0.7\%$, $1.2\%, 1.4\%, 0.5\%$ and $0.03\%$~\cite{ParticleDataGroup:2024}, respectively.

For each signal decay, the total systematic uncertainty in the BF measurement is obtained by adding all individual contributions in quadrature.
The results are summarized in Table~\ref{tab:totsys}.

\begin{table}[H]
	\centering
	\caption{Summary of the relative systematic uncertainties (\%) in the BF measurements. The ellipsis~($\cdots$) indicates that the corresponding source is not applicable.}
	\label{tab:totsys}
			\begin{tabular}{|c|c|c}
				\hline\hline
				\multicolumn{1}{|c|}{\multirow{1}{*}{Source}} &\multicolumn{1}{c|}{$K^{+} \pi^{-}\eta^\prime$}&
				\multicolumn{1}{c|}{$K^{+} \pi^{0}\eta^\prime$}  \\ \hline
				$e^-$ tracking &
				\multicolumn{1}{c|}{0.5}  &\multicolumn{1}{c|}{$\cdots$}\\
				$e^-$  PID &
				\multicolumn{1}{c|}{0.5}   &\multicolumn{1}{c|}{$\cdots$}\\
				$K^\pm$ or $\pi^\pm$ tracking &
				\multicolumn{1}{c|}{2.0}   &\multicolumn{1}{c|}{1.5}\\ 
				$K^\pm$ or $\pi^\pm$  PID &
				\multicolumn{1}{c|}{2.0}   &\multicolumn{1}{c|}{1.5}\\ 
				$K^\pm/\pi^\pm$ misidentification&
				\multicolumn{1}{c|}{0.5}   &\multicolumn{1}{c|}{0.5}\\ 
				$\eta(\pi^0)$ reconstruction &
				\multicolumn{1}{c|}{1.0}     &\multicolumn{1}{c|}{2.0}\\ 
				$\eta^\prime$ mass window &
				\multicolumn{1}{c|}{5.2}   &\multicolumn{1}{c|}{3.6} \\ 
				$\Delta E$ and $M^{\rm sig}_{\rm BC}$ requirements &
				\multicolumn{1}{c|}{0.3}   &\multicolumn{1}{c|}{0.6} \\ 
				$M_{K^+e^-}$ requirement &
				\multicolumn{1}{c|}{0.1}   &\multicolumn{1}{c|}{$\cdots$} \\
				$E^{\rm max}_{\rm extra\,\gamma}\&N^{\pi^0}_{\rm extra}\&N^{\rm track}_{\rm extra}$ &
				\multicolumn{1}{c|}{5.8}   &\multicolumn{1}{c|}{$\cdots$}\\
				$U_{\rm miss}$ and $M_{\rm BC}$ fit &  
				\multicolumn{1}{c|}{1.9}   &\multicolumn{1}{c|}{Neglected}\\
				MC statistics &
				\multicolumn{1}{c|}{0.2}  &\multicolumn{1}{c|}{0.2}\\ 
				MC modeling &
				\multicolumn{1}{c|}{1.9}   &\multicolumn{1}{c|}{1.4}\\ 
				FSR recovery &
				\multicolumn{1}{c|}{0.3}   &\multicolumn{1}{c|}{$\cdots$}\\
				$N_{D^0\bar D^0}$ &
				\multicolumn{1}{c|}{1.2}  & \multicolumn{1}{c|}{$\cdots$}\\ 
				$N_{\rm ST}$ &
				\multicolumn{1}{c|}{$\cdots$}    &\multicolumn{1}{c|}{0.3}\\ 
				Quoted BFs &
				\multicolumn{1}{c|}{1.5}   &\multicolumn{1}{c|}{1.5}\\  \hline
				Total   &
				\multicolumn{1}{c|}{9.0}   &\multicolumn{1}{c|}{5.1}\\ \hline
				\hline
			\end{tabular}
		\end{table}
	\begin{table}[H]
	\centering
	\caption{Fixed parameters used in the joint likelihood method.}
	\scalebox{0.9}{
		\begin{tabular}{|c|c|c|c|c|c|c|}
			\hline
			\hline
			Decay mode& $N^{\rm obs}$&$N_{\rm bkg1}$&$N_{\rm bkg2}^{\rm MC}$&$\sigma$ & $N^{\rm eff}(\times 10^{3})$ &$r$ \\
			\hline
			$D^0\to K^+\pi^-\eta^\prime$&6&4&3.0$\pm$0.3&0.09&184.0$\pm$2.1&0.879\\
			$D^+\to K^+\pi^0\eta^\prime$&1&0&1.3$\pm$0.2&0.05&148.6$\pm$0.1&1.094\\
			\hline
			\hline
		\end{tabular}
		\label{tab::jlf}
	}
\end{table}
\begin{table*}[htbp]
	\centering
	\caption{The BFs of $D^0\to K^+\pi^-\eta^\prime$ and $D^+\to K^+\pi^0\eta^\prime$ measured in this work~$(\mathcal{B}_{\rm DCS}^{\rm This work})$, the PDG values of the BFs of the corresponding CF decays~$(\mathcal{B}_{\rm CF}^{\rm PDG})$, the DCS/CF ratios~$(\mathcal{B}_{\rm DCS}^{\rm This work}/\mathcal{B}_{\rm CF}^{\rm PDG})$ and the BF ratio in unit of $\tan^{4}{\theta}_{C}$.}
	\renewcommand\arraystretch{1.3}
				\begin{tabular}{|c|c|c|c|c|c|}
					\hline
					\hline
					CF&$\mathcal{B}(\times 10^{-3})$& DCS& $\mathcal{B}(\times 10^{-5})$ &$\mathcal{B}_{\rm DCS}/\mathcal{B}_{\rm CF}$(\%)&$\mathcal{R}(\times \tan^4\theta_C)$ \\
					\hline
					$D^0\to K^-\pi^+\eta^\prime$&6.43$\pm$0.34&$D^0\to K^+\pi^-\eta^\prime$ &$<3.0$ &$<0.5$&$<1.6$ \\
					$D^+\to K_S^0\pi^+\eta^\prime$&1.90$\pm$0.21&$D^+\to K^+\pi^0\eta^\prime$ &$<2.1$ &$<1.1$&$<3.7$ \\
					\hline
					\hline
				\end{tabular}
			\label{totresult}
		\end{table*}

\section{Results}
\begin{figure*}[htbp]
	\centering
	\includegraphics[width=0.45\linewidth]{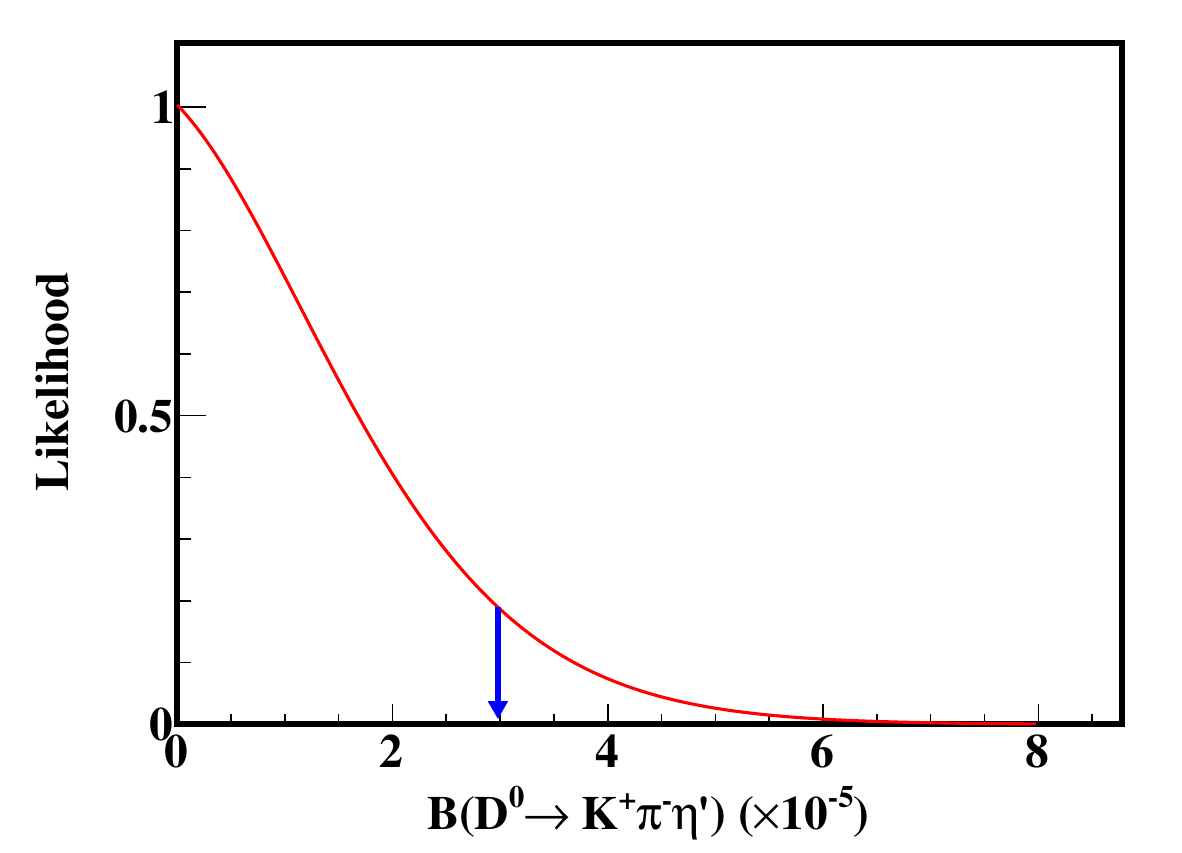}
	\includegraphics[width=0.45\linewidth]{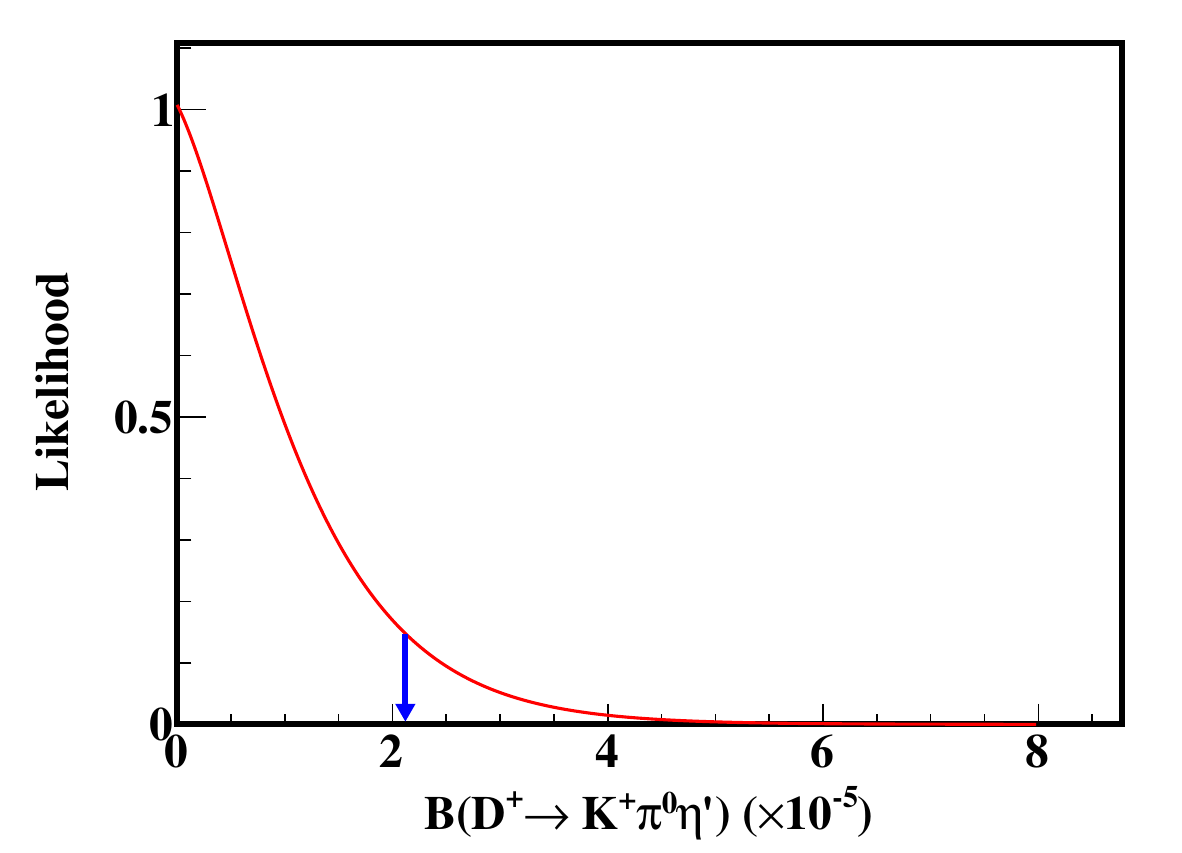}
	\vspace{6pt}
	\caption{Likelihood curves for the BFs of the signal  decays $D^0\to K^+\pi^-\eta^\prime$ and $D^+\to K^+\pi^0\eta^\prime$. The red curves denote the likelihood results. The blue arrows denote the upper limits at the 90\% confidence level. }
	\label{fig::jlm}
\end{figure*}

The upper limits on the BFs are determined using an extended profile likelihood method
~\cite{PLM}. According to Eq.~\ref{equ:br} and~\ref{eq:br}, the number of signal events $N_{\rm sig}$ satisfies $N_{\rm sig} = N^{\rm eff}\mathcal{B}_{\rm sig}$. For $D^0\to K^+\pi^-\eta^\prime$, $N^{\rm eff} = 2\cdot N_{\rm D^0\bar{D}^0}\cdot \epsilon_{\rm DT}\cdot \mathcal{B}^{\rm inter}$. For $D^+\to K^+\pi^0\eta^\prime$, $N^{\rm eff} = N^{\rm ST}\cdot \epsilon_{\rm sig} \cdot \mathcal{B}^{\rm inter}$. Here, $\mathcal{B}^{\rm inter}$ is the product of the intermediate BFs such as $\mathcal{B}(\pi^0\to\gamma\gamma)$. For $N^{\rm eff}$, a Gaussian distribution is assumed, with the mean equal to the measured value of $N^{\rm eff}$ and the standard deviation equal to its absolute systematic uncertainty, $N^{\rm eff}\cdot \sigma$, where $\sigma$ represents the relative total systematic uncertainty listed in Table~\ref{tab:totsys}. The joint likelihood function~\cite{jlm} is given by
\begin{flalign}
	\mathcal{L}&=\mathcal{P}(N^{\rm obs}, N^{\rm eff}\cdot \mathcal B+N_{\rm bkg1} + N_{\rm bkg2}^{\rm MC}) \\ \nonumber
	&\cdot\mathcal{G}(N^{\rm eff},N^{\rm eff},N^{\rm eff}\cdot \sigma) \\ \nonumber
	&\cdot \mathcal{P}(N_{\rm bkg1}, N_{\rm bkg1}/r) \\ 
	&\cdot \mathcal{G}(N_{\rm bkg2}^{\rm MC}, N_{\rm bkg2}^{\rm MC},\sigma_{\rm bkg2}^{\rm MC}), \nonumber 
\end{flalign}
where $N^{\rm obs} = N_{\rm sig}+N_{\rm bkg1}+N_{\rm bkg2}^{\rm MC}$ and $N^{\rm eff} = \mathcal{B}^{\rm inter}\cdot N^{\rm ST}\cdot \epsilon^{\rm sig}$; $N_{\rm bkg1}$ is the number of background events originating from misreconstructed tag $\bar{D}$ candidates, which is estimated by data; $N_{\rm bkg2}^{\rm MC}$ is the number of background events originated from correctly reconstructed $\bar{D}$ and misreconstructed signal $D$ candidates, which is estimated by inclusive MC sample;
$r$ is the ratio of background events in the $M_{\rm BC}^{\rm tag}$ signal and sideband regions in data, estimated using the hadronic tag for $D^0\to K^+\pi^-\eta^\prime$ and the $D^-\to K^+\pi^-\pi^-$ tag mode for $D^+\to K^+\pi^0\eta^\prime$; $\sigma$ denotes the total systematic uncertainty. The fixed parameters for the $D^0\to K^+\pi^-\eta^\prime$ and $D^+\to K^+\pi^0\eta^\prime$ modes are listed in Table~\ref{tab::jlf}. With this function, we scan the $\mathcal{B}$ from 0 to $9\times 10^{-5}$ to obtain the upper limits at the 90\% confidence level; the corresponding likelihood curves are shown in Fig.~\ref{fig::jlm}.

\section{Summary}
Based on an analysis of $e^+e^-$ collision data collected at $\sqrt{s} = 3.773$~GeV with
the BESIII detector, corresponding to an integrated luminosity of 20.3~fb$^{-1}$,
we search for the DCS decays $D^0\to K^+\pi^-\eta^\prime$ and $D^+\to K^+\pi^0\eta^\prime$ for the first time. The upper limits on their BFs are measured to be 
\begin{equation}
	\mathcal{B}(D^0\to K^+\pi^-\eta^\prime)<3.0\times10^{-5} \nonumber
\end{equation}
and 
\begin{equation}
	\mathcal{B}(D^+\to K^+\pi^0\eta^\prime)<2.1\times 10^{-5}\nonumber
\end{equation}
 at 90\% confidence level, respectively.
Table~\ref{totresult} summarizes the BFs measured in this work, the PDG values of the BFs of the corresponding  CF decays, the individual DCS/CF ratios and the factors in unit of $\tan^{4}{\theta}_{C}$.
At present, no definite conclusion can be made for these two DCS decays. These two decays are expected to be observed with the much larger data sample to be collected at the Super Tau-Charm Facility~\cite{STCF}.

\section{Acknowledgement}
The BESIII Collaboration thanks the staff of BEPCII (https://cstr.cn/31109.02.BEPC) and the IHEP computing center for their strong support. This work is supported in part by National Key R\&D Program of China under Contracts Nos. 2023YFA1606000, 2023YFA1606704, 2025YFA1613900; National Natural Science Foundation of China (NSFC) under Contracts Nos. 12635006, 11635010, 11935015, 11935016, 11935018, 12025502, 12035009, 12035013, 12061131003, 12192260, 12192261, 12192262, 12192263, 12192264, 12192265, 12221005, 12225509, 12235017, 12342502, 12361141819, 12535005; the Chinese Academy of Sciences (CAS) Large-Scale Scientific Facility Program; the Strategic Priority Research Program of Chinese Academy of Sciences under Contract No. XDA0480600; CAS under Contract No. YSBR-101; 100 Talents Program of CAS; The Institute of Nuclear and Particle Physics (INPAC) and Shanghai Key Laboratory for Particle Physics and Cosmology; ERC under Contract No. 758462; German Research Foundation DFG under Contract No. FOR5327; Istituto Nazionale di Fisica Nucleare, Italy; Knut and Alice Wallenberg Foundation under Contracts Nos. 2021.0174, 2021.0299, 2023.0315; Ministry of Development of Turkey under Contract No. DPT2006K-120470; National Research Foundation of Korea under Contract No. NRF-2022R1A2C1092335; National Science and Technology fund of Mongolia; Polish National Science Centre under Contract No. 2024/53/B/ST2/00975; STFC (United Kingdom); Swedish Research Council under Contract No. 2019.04595; U. S. Department of Energy under Contract No. DE-FG02-05ER41374

\onecolumngrid

\begin{center}
  \rule{3cm}{1.5pt}
\end{center}
\end{document}